\documentclass{optica-article}

\journal{opticajournal} % for journals or Optica Open

\articletype{Research Article}

\usepackage{lineno}

\hypersetup{
    colorlinks=true,
    citecolor=blue,  % make \cite{} links blue
    linkcolor=blue, % keep figure/table refs black
    urlcolor=blue    % make URLs blue (optional)
}
\begin{document}

\title{Voltage-controlled beam steering in liquid-crystal-integrated dual-mode plasmonic nanolaser}

\author{Md.~Rasel Parvez\authormark{1} and Muhammad Anisuzzaman Talukder\authormark{1,*}}

\address{\authormark{1}Department of Electrical and Electronic Engineering, Bangladesh University of Engineering and Technology, Bangladesh\\
}

\email{\authormark{*}anis@eee.buet.ac.bd} %% email address is required; see note below about the corresponding author designation

% use {asbstract*} to suppress the copyright line. Copyright information will be added in production

\begin{abstract*} 
Dynamic control of laser emission direction is crucial for developing compact and reconfigurable nanophotonic devices. In this work, we numerically present a plasmonic nanolaser (PNL) integrated with a voltage-controlled liquid crystal (LC) layer to achieve active beam steering. We modeled the orientation of LC molecules under an applied bias and incorporated this into electromagnetic simulations to assess the optical response. We observed lasing at an emission wavelength of 870 nm for the single-mode PNL, with discrete voltage-dependent deflections of the far-field emission of up to $\pm67$\textdegree, while maintaining a beam divergence of less than 1\textdegree. The steering characteristics were significantly influenced by the electro-optic properties of the LC layer, with an optimized thickness of 3 $\mu$m. The structural periodicity governed the achievable angular separation and emission stability. Furthermore, we extended the concept to a dual-mode nanolaser based on a merged nanohole array (NHA), which supports two lasing wavelengths at 873 nm and 880 nm. Both lasing modes exhibited simultaneous voltage-dependent angular tuning without altering the cavity geometry. These results highlight the potential of LC-integrated PNLs as voltage-controlled, reconfigurable light sources for applications in optical interconnects, beam routing, adaptive imaging, and multiplexed communication.
\end{abstract*}

%%%%%%%%%%%%%%%%%%%%%%%%%%  body  %%%%%%%%%%%%%%%%%%%%%%%%%%
\section{Introduction}

At the nanoscale, steering the direction of light on demand is essential for nanophotonic platforms, including adaptive photonic circuits, on-chip optical interconnects, reconfigurable sensors, and nanoscale imaging systems \cite{koo2002bragg,momtaj2025toward,pleros2011optical,stucchi2013chip,ohashi2009chip,ozcelik2015optofluidic,zia2006plasmonics}. Mechanical mirrors, MEMS, and phased arrays are effective at larger scales; however, they encounter limitations in terms of footprint, speed, integration with nanoscale emitters, and overall system overhead. This gap motivates the development of compact, fast, and electrically tunable steering methods that pair naturally with plasmonic nanolasers (PNLs), which offer subwavelength confinement and strong field localization.

PNLs are strong candidates for coherent emission at the nanoscale because they squeeze optical energy below the diffraction limit through excitation of surface plasmon modes \cite{bergman2003surface,hill2007lasing,ma2011room,noginov2009demonstration,lu2012plasmonic,khajavikhan2012thresholdless,deeb2017plasmon,zhang2014room,azad2022}. The architecture enables intense interaction between light and matter within a tiny footprint, large Purcell enhancement and low mode volume \cite{hill2007lasing}, and the prospect of ultrafast operation \cite{ma2011room}. Even so, most conventional PNLs emit at angles set by the cavity geometry, and real-time control of the far field remains difficult. Multimode designs are attractive for multi-target sensing, interferometry, and multiplexed links, but they often introduce off-normal emission, broader angular patterns, and higher thresholds due to symmetry breaking or competition between modes \cite{zumrat2022dual,azad2021mode,ahmed2018efficient,shahid2025beyond,ahamed2024wavelength,Winkler2020DualWavelengthLI}.

Several previous works have established the foundation for this study by progressively enhancing PNL architectures developed within our research group. In 2018, Ahmed and Talukder \cite{ahmed2018efficient} demonstrated an efficient and highly directional single-mode PNL based on the coupling of optical Tamm states with a distributed Bragg reflector and a metal NHA. This structure provided a compact cavity with strong vertical confinement and narrow beam divergence, forming the baseline for subsequent designs. Building on this concept, Zumrat and Talukder \cite{zumrat2022dual} introduced a merged NHA configuration that broke the in-plane symmetry of the lattice, enabling dual-mode lasing at distinct wavelengths from a single device. More recently, Ahamed et al. \cite{ahamed2024wavelength} proposed a metasurface-integrated dual-mode PNL capable of wavelength-selective beam steering, where a dielectric phase-gradient metasurface spatially separated the two lasing modes. The merged lattice dual mode PNL of Zumrat and Talukder \cite{zumrat2022dual} provided two lasing modes from a single cavity. However, the beam direction was fixed by the lattice geometry and not dynamically controlled. Similarly, the metasurface-integrated dual-mode PNL of Ahamed et al.~\cite{ahamed2024wavelength} achieved wavelength-selective beam separation, but the steering angles were locked by the static phase gradient metasurface. In both cases, the emission direction would essentially be set at design and would not be possible to change once the devices are fabricated.

Liquid crystals (LCs) provide a practical route to dynamically control light propagation in plasmonic platforms \cite{mur2022controllable}. Because the director angle follows an applied voltage, the local refractive index changes continuously, which lets beams steer without mechanical movement \cite{yang2014fundamentals,bahadur1990dynamic,lu2005ultrawide,assanto2012nematicons,sautter2015active,komar2018tunable}. LC layers have already tuned photonic crystals, diffraction gratings, and metasurfaces. However, pairing LCs with PNLs for dynamic beam steering, especially in dual-mode operation, is an unexplored area. Combining voltage-controlled LC orientation with PNLs suggests a compact, low-power, and reconfigurable method for setting emission direction and spectral content in real-time.

This work designs and analyzes the dynamic steering of PNL emission using an integrated LC layer. Applying a bias across the LC cell rotates the director; the resulting index map sets the laser’s emission angle. We compute LC profiles in COMSOL Multiphysics for multiple voltages, then import them into Lumerical FDTD to simulate single-mode and dual-mode nanolaser layouts \cite{lumericalFDTD,COMSOL62}. The analysis spans voltage-dependent orientation, emission spectra, far-field intensity maps, as well as the roles of LC thickness and periodicity in setting the steering range and stability. To the best of our knowledge, this is the first numerical demonstration of voltage-controlled dynamic beam steering in plasmonic nanolasers, and the first dual-mode PNL in which both lasing modes can be steered electrically without modifying the cavity or metasurface geometry.

With the LC layer in place, emission direction is actively driven at the nanoscale, providing an electrically tunable alternative to static metasurfaces or fixed cavity geometries. In our implementation, a single-mode PNL at 870 nm is redirected over a $\pm$67\textdegree angular range, while the beam divergence remains $<1$\textdegree. The steering response depends on LC parameters. A 3-$\mu$m thickness gives the best balance between output strength and angular stability. Extending the approach to a merged lattice NHA yields two lasing modes at 873 nm and 880 nm, whose emission angles shift with the applied voltage while maintaining their spectral positions intact. Unlike previous metasurface- or merged-lattice-based designs that fixed emission direction at fabrication, this work introduces post-fabrication, voltage-programmable angular tuning of both lasing modes within a single cavity, offering a reconfigurable functionality not previously achieved. These results outline a practical route to compact, voltage-programmable nanophotonic emitters that support beam steering, adaptive coupling, and on-chip optical links within a single platform.

\section{Device Configuration and Simulation Methodology}
\subsection{Device Architecture and Structural Design}
The PNL considered in this work builds on a previously reported design by Ahmed et al.~\cite{ahmed2018efficient}, which integrates a metallic NHA on a thin Au film, a dye-doped polyurethane (PU) gain layer, and a one-dimensional photonic crystal reflector. The perforated gold supplies the plasmonic feedback required for lasing. Optical gain is provided by PU loaded with IR-140 dye. Beneath the gain medium, alternating TiO$_2$ and MgF$_2$ layers form a distributed Bragg reflector that strengthens vertical confinement of the lasing mode. This hybrid scheme has been widely investigated in PNL research for its ability to combine strong confinement with efficient outcoupling of coherent emission~\cite{azad2021mode,zumrat2022dual,shahid2025beyond}.
%
%---------------------------------
% Figure 1
\begin{figure*}[t]
\centering
\includegraphics[width=\textwidth, clip=true, trim=0cm 16cm 3.75cm 0cm]{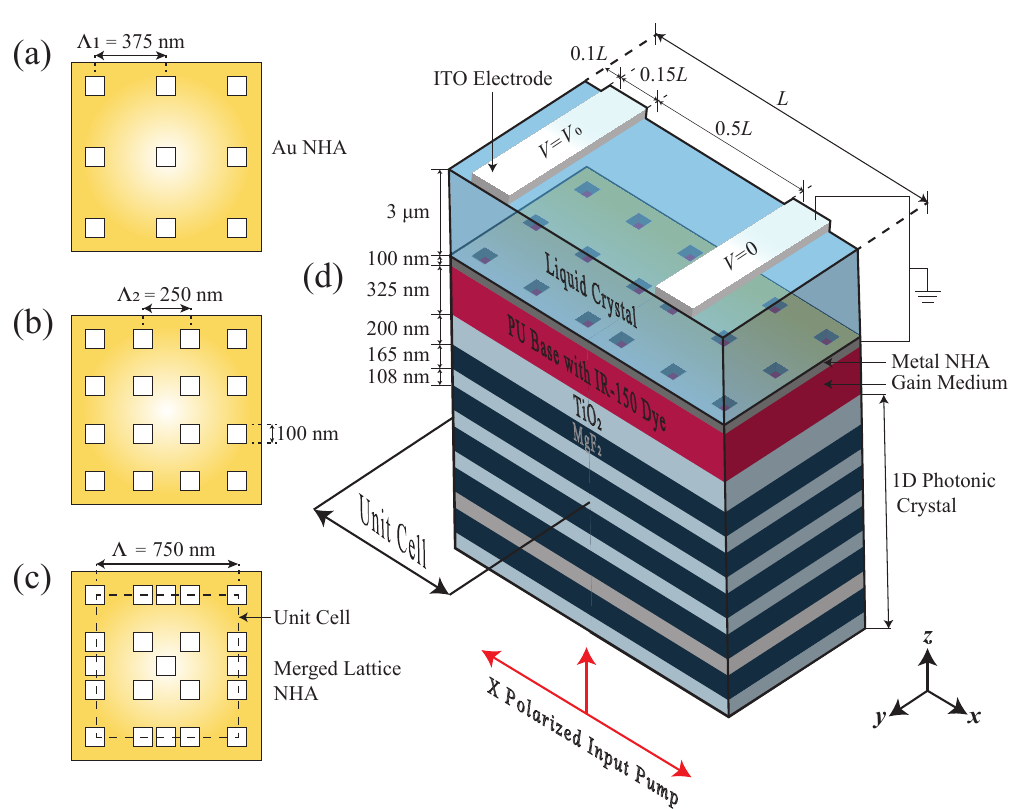}
\caption{(a) Top view of the NHA with unit cell period $\Lambda_{1} = 375$ nm used in the single mode plasmonic nanolaser, (b) Top view of the NHA with unit cell period $\Lambda_{2} = 250$ nm, (c) Top view of merged lattice with unit cell period $\Lambda = 750$ nm employed in the dual-mode plasmonic nanolaser, (d) Three-dimensional schematic of the complete device showing the LC layer and patterned ITO electrodes integrated above the metallic NHA, gain medium, and TiO$_2$/MgF$_2$ photonic crystal stack. Applying a voltage across the electrodes reorients the LC molecules, enabling dynamic beam steering of the emitted laser. The structural dimensions and layer thicknesses are not drawn to scale and are shown schematically for clarity.}
\label{fig:1}
\end{figure*}
%----------------------------------------

In the single-mode configuration, the structure consisted of a square NHA with a period of $\Lambda_{1}=375$ nm, as shown in Fig.~\ref{fig:1}(a), patterned on a 100 nm-thick gold film in the earlier design \cite{ahmed2018efficient}. This geometry supports a single plasmonic mode that couples efficiently to the dye emission around 870 nm. The gain medium consists of a 325 nm-thick PU layer doped with IR-140 at a concentration of $2 \times 10^{18}$ cm$^{-3}$, exhibiting an absorption peak near 800 nm and an emission peak at 870 nm, making it suitable for near-infrared lasing applications.

To achieve dual-mode operation, a merged NHA design was utilized, as shown in Fig.~\ref{fig:1}(c), in which two lattices with periods $\Lambda_{1}=375$ nm (Fig.~\ref{fig:1}a) and $\Lambda_{2}=250$ nm (Fig.~\ref{fig:1}b) were superimposed. This merged lattice supports two distinct plasmonic modes within the same cavity, enabling simultaneous dual wavelength lasing, as reported in previous studies on multi-mode PNLs \cite{zumrat2022dual}. The dual-mode structure thus serves as an effective platform for exploring voltage-dependent beam steering through LC integration.
This integrated design with its single and dual mode variants forms the baseline platform for the LC-enabled beam steering studies reported in this work.

To enable dynamic beam steering, a nematic LC cell is integrated on top of the PNL. The device layout is shown in Fig.~\ref{fig:1}(d). A $3~\mu$m-thick LC layer is placed above the metallic NHA, which simultaneously functions as the optical feedback layer and the ground electrode. Transparent indium tin oxide (ITO) electrodes are deposited on the top surface of the LC cell. All lateral dimensions are expressed relative to $L$, the effective length of the LC cell as defined in Eq.~\eqref{eq:Ldef}. Each electrode has a width of $0.15L$, and the two electrodes are separated by a central gap of $0.5L$, which defines the active modulation region. This electrode geometry ensures that the applied electric field is confined primarily within the LC layer above the laser aperture, minimizing fringing effects at the edges. The effective length of the LC cell is given by  
%
%-----------------
\begin{equation}
L = (2N+1)\Lambda_{1} .
\label{eq:Ldef}
\end{equation}
%----------------

Here, $\Lambda_{1}$ is the NHA period shown in Fig.~\ref{fig:1}(a), and $N$ is the number of unit cells covered by a single LC region. Thus, $L$ can be interpreted as both the physical length of the LC cell and the periodicity of the LC structure. Since periodic boundary conditions are applied along the $x$- and $y$-axes in the optical simulations, $L$ also defines the unit periodic length of the overall device. For example, in Fig.~\ref{fig:1}(d), $N=2$, giving $L = 5\Lambda_{1}$. In our scheme, one of the two top ITO electrodes is driven while the other top electrode and the metal NHA beneath the LC are held at ground, as shown in Fig.~\ref{fig:1}(d). The LC molecules reorient along the electric field due to dielectric anisotropy. This reorientation alters the effective refractive index above the nanolaser, thereby modifying the phase profile of the emitted light and enabling dynamic control of the emission angle.  

The nematic compound 4-Cyano-4$'$-pentylbiphenyl (5CB) is used as the active LC material because of its high birefringence ($n_0 = 1.544$, $n_e = 1.736$), strong dielectric anisotropy, and stable electro-optic response  \cite{cirtoaje2020freedericksz}. The elastic constants ($K_1 = 6.2$ pN, $K_2 = 3.9$ pN, $K_3 = 8.2$ pN) ensure low threshold voltages for molecular reorientation in the chosen geometry, allowing efficient steering under modest bias.

\subsection{LC Orientation Modeling}
% LC (5CB) reorientation modeling — ready to paste
The reorientation of the nematic LC (5CB) under an applied bias was modeled in COMSOL Multiphysics. Two ITO electrodes were patterned on the top surface of the LC. During operation, one pad is driven, while the other pad and the metal NHA at the bottom are tied to ground, as shown in Fig.~\ref{fig:1}(d). This layout establishes a predominantly in-plane electric field across the LC, with weak vertical fringing near the driven ITO edge and toward the grounded metal NHA, which drives molecular reorientation in the active region.

The spatial director field $\mathbf{n}(x,y,z)$ was obtained by minimizing the total free energy of the nematic under the combined elastic and electric interactions. In a nematic, the long molecular axes align along a local unit vector $\mathbf{n}$ with $|\mathbf{n}|=1$. Once a field is applied, two torques compete: the elastic contribution resists deformation of $\mathbf{n}$, whereas the dielectric contribution tends to align $\mathbf{n}$ with $\mathbf{E}$. Equilibrium corresponds to the configuration that minimizes the total free energy density.

The total free energy density $f$ of a nematic LC follows the Frank Oseen form, which includes splay, twist, and bend deformations together with the dielectric term from the applied field \cite{frank1958liquid}
%--------------------
\begin{equation}
f = \frac{1}{2}\left[ K_{11} (\nabla \cdot \mathbf{n})^2 
+ K_{22} \big(\mathbf{n}\cdot\nabla\times\mathbf{n}\big)^2 
+ K_{33} \big(\mathbf{n}\times\nabla\times\mathbf{n}\big)^2 \right] 
- \frac{1}{2}\epsilon_0 \Delta \epsilon \big(\mathbf{n}\cdot \mathbf{E}\big)^2.
\label{eq:2}
\end{equation}
%---------------------------
%
Here $K_{11}$, $K_{22}$, and $K_{33}$ are the splay, twist, and bend elastic constants, which quantify the cost of each deformation. The first three terms of the right side of Eq.~(\ref{eq:2}) penalize spatial variation of $\mathbf{n}$, and the final term captures dielectric coupling, where $\Delta\epsilon=\epsilon_{\parallel}-\epsilon_{\perp}$ is the dielectric anisotropy. For positive $\Delta\epsilon$ (as in 5CB), the dielectric energy is the lowest when $\mathbf{n}\parallel\mathbf{E}$. When the applied voltage exceeds the Freedericksz threshold, $V_c\simeq 0.8~\mathrm{V}$, the dielectric torque outweighs the elastic torque. Consequently, the directors tilt toward the electric field, resulting in a corresponding change in the effective refractive index map within the LC layer. The resulting orientation profiles, computed as a function of applied voltage, were subsequently imported into the optical model of the PNL.

In COMSOL Multiphysics, the energy functional given in Eq.~(\ref{eq:2}) was implemented using the weak form PDE interface. The total free energy $F = \int f\, dV$ was minimized numerically with respect to two angular variables, $\theta(x,y,z)$ and $\phi(x,y,z)$, which parameterize the unit vector $\mathbf{n} = (\sin\theta\cos\phi, \sin\theta\sin\phi, \cos\theta)$. Solving the corresponding Euler--Lagrange equations in conjunction with the electrostatic potential yielded the steady-state director configuration for each applied voltage. This self-consistent approach captured the voltage-dependent molecular reorientation that was later used as input for the optical simulation of the PNL.

Strong anchoring conditions, implemented as Dirichlet boundary constraints on the director field, were applied at the top and bottom surfaces of the LC cell to enforce planar alignment in the absence of bias. Periodic boundary conditions were set along the $x$- and $y$-axes to reflect the periodic nature of the NHA. The electrode geometry consisted of two ITO strips of width $0.15L$ separated by a central gap of $0.5L$, where $L = (2N+1)\Lambda_{1}$ denotes the length of one LC simulation cell and $\Lambda_{1}$ is the period of the NHA unit cell. The material parameters of 5CB used in the simulations are summarized in Table~\ref{tab:5cb}.  

%----------------------------------
% Table 1
\begin{table}[htbp]
\centering
\caption{Material parameters for nematic liquid crystal 5CB \cite{cirtoaje2020freedericksz}.}
\label{tab:5cb}
\begin{tabular}{ll}
\hline
Parameter & Value \\
\hline
Anisotropic susceptibility, $\chi_a$ & $1.46\times10^{-6}$ \\
Splay elastic constant, $K_{11}$ & 6.2 pN \\
Twist elastic constant, $K_{22}$ & 3.9 pN \\
Bend elastic constant, $K_{33}$ & 8.2 pN \\
Ordinary refractive index, $n_0$ & 1.544 \\
Extraordinary refractive index, $n_e$ & 1.736 \\
\hline
\end{tabular}
\end{table}
%------------------------------

\subsection{Optical Simulation Methodology}
The optical response of the LC-integrated PNL was investigated using the finite-difference time-domain method implemented in Lumerical FDTD Solutions \cite{lumericalFDTD}. Three-dimensional full-field vectorial Maxwell’s equations coupled with rate equations of the gain medium were solved to capture the lasing dynamics \cite{yee1966numerical}. The simulation domain consisted of the PNL structure together with the integrated LC cell and top ITO electrodes. Periodic boundary conditions were applied along the $x$- and $y$-axes to represent the infinite array, while perfectly matched layer boundaries were used along the $z$-axis to absorb outgoing radiation. A non-uniform conformal meshing scheme was adopted, with a minimum mesh size of 0.25 nm in the vicinity of the NHA to accurately resolve the plasmonic features.  

The gain medium was modeled as IR-140 dye molecules embedded in a PU host medium using a four-level two-electron atomic system \cite{dridi2013model,chang2004finite}. In this four-level two-electron system, an \(800\,\mathrm{nm}\) optical pump excites the carriers from level 0 to 3. Fast nonradiative relaxation transfers population from level 3 to 2 and establishes inversion. Stimulated emission occurs on the transition from level 2 to 1 at $\sim$\(870\,\mathrm{nm}\). A rapid decay from level 1 to 0 returns the system to the ground state. This model captures the essential population dynamics through coupled rate equations, without explicitly tracking every transition. The PU host has a refractive index of 1.51 and is doped with IR-140 dye at a concentration of $2 \times 10^{18}\,\text{cm}^{-3}$. The dye molecules provide optical gain via stimulated emission around 870 nm, with absorption centered at 800 nm. The key parameters used in the gain medium model are summarized in Table~\ref{tab:gain}.  

%--------------------------
\begin{table}[htbp]
\centering
\caption{Parameter values of the four-level two-electron gain medium model}
\label{tab:gain}
\begin{tabular}{lc}
\hline
\textbf{Parameter} & \textbf{Value} \\
\hline
Emission wavelength, $\lambda_a$ & 870 nm \\
Emission linewidth, $\Delta\lambda_a$ & 100 nm \\
Absorption wavelength, $\lambda_b$ & 800 nm \\
Absorption linewidth, $\Delta\lambda_b$ & 100 nm \\
Gain medium base material index & 1.51 \\
Dye concentration & $2 \times 10^{18}$ cm$^{-3}$ \\
Polarization decay rate, $\gamma_a=\gamma_b$ & $3.9 \times 10^{13}$ s$^{-1}$ \\
Radiative decay rate, $\gamma_{\text{rad}}$ & $7.2 \times 10^{7}$ s$^{-1}$ \\
Transition lifetimes: & \\
\quad $\tau_{30} = \tau_{21}$ & 1 ns \\
\quad $\tau_{32} = \tau_{10}$ & 10 fs \\
\hline
\end{tabular}
\end{table}
%--------------------------------

The device was optically pumped using a plane wave source with a central wavelength of 800 nm, pulse width of 40 fs, and electric field polarized along the $x$-direction. The pump was injected from the $-z$ side and propagated toward the $+z$ direction through the DBR into the gain medium. The lasing emission exited through the metallic NHA in the $+z$ direction.

Near-field monitors inside the cavity and far-field projection monitors above the LC cell were used to analyze the lasing spectra and emission profiles. Both single-mode ($\sim870$ nm) and dual-mode ($\sim873$ nm and $\sim880$ nm) lasing cases were investigated. By sweeping the applied voltage across the LC cell, the anisotropic permittivity distribution obtained from COMSOL was incorporated into the FDTD model, allowing for the direct evaluation of the beam steering characteristics of the integrated nanolaser.

\subsection{Studied Parameters}
To characterize beam steering in the LC integrated PNL, three control parameters were investigated: the drive voltage applied to the top ITO pads, the LC thickness, and the effective periodicity of the LC cell. For each voltage, we exported the LC director map from COMSOL, along with the far-field emission profile from Lumerical FDTD Solutions. Pairing these datasets reveals that the voltage-driven reorientation of the LC results in systematic shifts of the nanolaser emission angle in the far field.

Changing the LC thickness reveals the tradeoff between phase delay and output power. Thin layers yield small index modulation, so steering weakens. Thick layers increase phase retardation but reduce the emitted intensity because absorption and scattering rise. By comparing the far-field and spectral responses for different thickness values, an optimal configuration was identified.

The role of periodicity was examined by varying the number of nanolaser unit cells covered by a single LC segment. The effective length of one LC cell is given by $L=(2N+1)\Lambda_{1}$, where $\Lambda_{1}$ is the NHA period. Sweeping $N$ changes the set of accessible steering angles and their spacing.

\section{Results and Discussion}
\subsection{LC Orientation Profiles under Applied Electric Field}

In our simulation, the LC layer is placed between two patterned ITO pads on the top side and the gold NHA at the bottom, with the NHA serving both as the optical interface and the common electrode. The NHA is held at ground, and the upper ITO pads act as control electrodes. A voltage is applied to the left pad, while the right pad is kept at ground, so that an in-plane electric field is established across the LC region, and pronounced lateral fringing fields appear between the biased and grounded sides.

%------------------------------------
% Figure 2
\begin{figure*}[htbp]
\centering
\includegraphics[width=\textwidth, clip=true, trim=0cm 17.75cm 0cm 0cm]{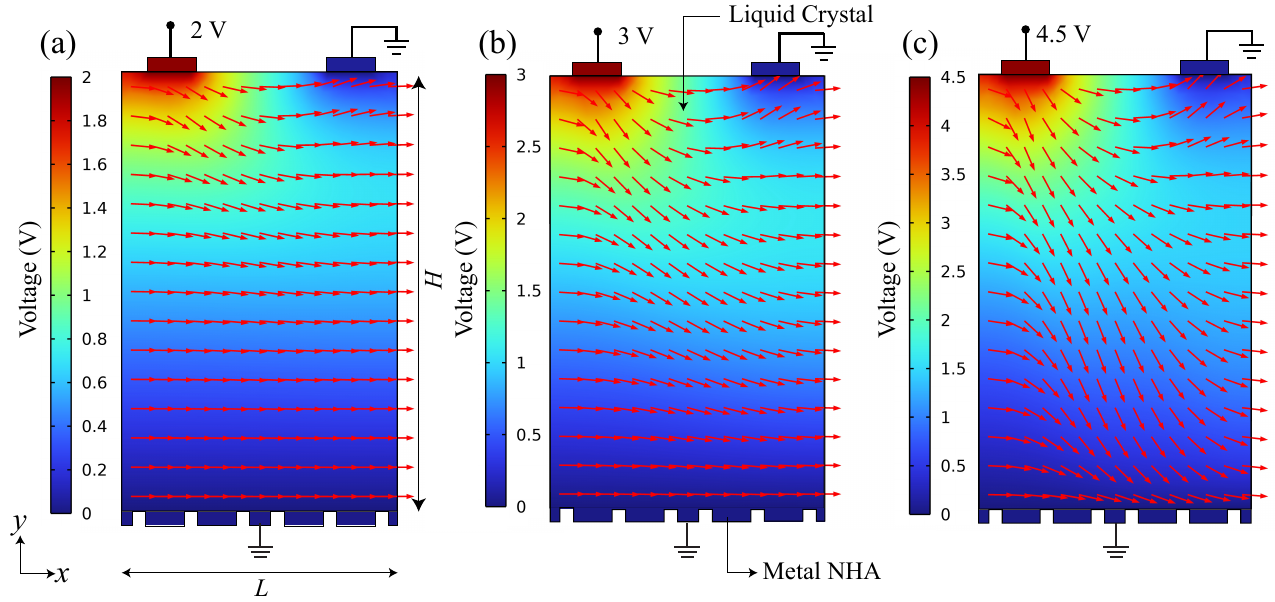}
\caption{LC orientation profiles under an applied electric field.
The color map indicates the potential distribution inside the nematic LC cell,
and the arrows show the local director orientation $\mathbf{n}$ for 
(a) $V = 2$~V, (b) $V = 3$~V, and (c) $V = 4.5$~V applied across the ITO electrodes.}
\label{fig:LC_orientation}
\end{figure*}
%-------------------------------------

Strong anchoring is imposed at the top and bottom LC interfaces so that the directors remain fixed at the boundaries, while periodic boundary conditions are applied along the lateral directions to represent an extended array. At low bias, the director field remains close to the initial uniform alignment, with only mild distortion near the driven electrode. As the applied voltage increases, reorientation is no longer confined to the immediate vicinity of the driven electrode. LC molecules near the biased ITO pad follow the local field lines and tilt more strongly, while those closer to the grounded side respond more weakly.

This non-uniform director field introduces a smooth variation in optical phase across the emitting aperture. At higher bias, the phase difference between opposite sides of the cell grows, and the LC layer effectively behaves as a controllable phase element for the outgoing beam. The profiles in Fig.~\ref{fig:LC_orientation} illustrate this trend: at $V = 2$~V the distortion is localized near the biased pad, at $V = 3$~V it extends further into the bulk, and at $V = 4.5$~V a large portion of the cell is tilted, which is sufficient to produce a clear deflection in the far-field emission angle.

\subsection{Single-Mode Nanolaser: Beam Steering Characteristics}

\begin{figure*}[t]
\centering
\includegraphics[width=\textwidth, clip=true, trim=0cm 20.6cm 0cm 0cm]{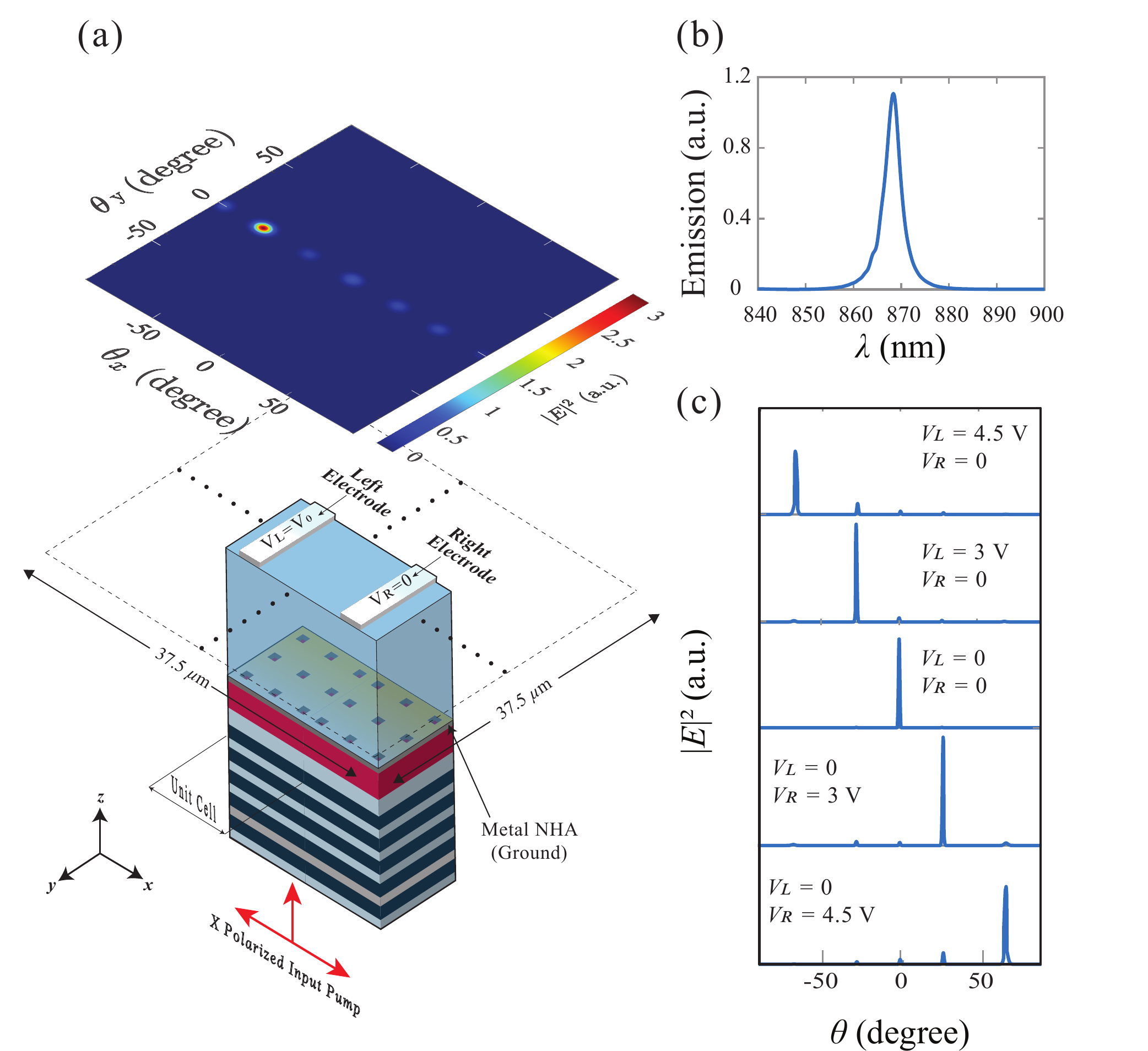}
\caption{Beam steering of the single-mode plasmonic nanolaser. 
(a) Integrated device structure with an LC cell placed on top of the PNL and two patterned ITO electrodes for applying asymmetric bias. 
(b) Emission spectrum showing a narrow lasing peak at 870~nm with a linewidth of 2~nm. 
(c) Far-field intensity distribution under different applied voltages, illustrating the transition of dominant emission from the zeroth-order beam at 0–2~V to the first-order mode at $-27^{\circ}$ (3~V), and finally to the second-order mode at $-67^{\circ}$ (4.5~V).}
\label{fig:SM_beam}
\end{figure*}

The steering behavior of the single-mode PNL was investigated by integrating an LC cell on top of the cavity and applying asymmetric voltages to the patterned ITO electrodes. The Au NHA acts as the bottom electrode and simultaneously provides the plasmonic feedback necessary to sustain lasing, while the bias is introduced through either the left or right top electrode. For the configuration studied here, the LC covers $N=2$ unit cells of the NHA, with the device periodic along both $x$- and $y$-directions. The simulation domain spans an effective size of $37.5~\mu$m $\times$ $37.5~\mu$m, and the structure is pumped with an $x$-polarized plane wave of wavelength 800~nm at an intensity of $2\times10^{8}$~V/m, as shown in Fig.~\ref{fig:SM_beam}(a).  

%
%-----------------------------------------
% Figure 3
%--------------------------------------

The device exhibits a narrow lasing line near 870 nm with a 2 nm linewidth, confirming stable single-mode operation of the LC integrated nanolaser, as shown in Fig.~\ref{fig:SM_beam}(b). The spectrum remains essentially unchanged as the bias is varied, so the observed steering originates from LC-driven phase modulation rather than multimode competition or spectral broadening. The emission intensity at the steered angle increases with applied bias, consistent with a stronger LC index gradient directing a larger fraction of the power into that radiative order while the lasing wavelength remains fixed.

The far field evolves with bias in a voltage-dependent redistribution among diffraction orders, as shown in Fig.~\ref{fig:SM_beam}(c). At 0~V, the emission is dominated by the zeroth-order beam along the surface normal, a condition that remains stable up to 2~V. However, beyond 2 V, the first-order mode at $-27^\circ$ begins to grow steadily and, by 3~V, it becomes the dominant mode, causing the main beam to be deflected away from the normal position. As the voltage increases, higher-order modes are further enhanced. At 4.5 V, the second-order mode at $-67^\circ$ takes the lead. Intermediate voltage levels result in multiple angular peaks, which are consistent with partial LC reorientation and a nonuniform phase gradient across the aperture. Across the entire sweep, each angular lobe maintains a full width at half maximum of $\sim1^\circ$, indicating preserved beam quality. The steering direction follows the driven electrode: biasing the left pad yields $\theta<0^\circ$, whereas driving the right pad gives $\theta>0^\circ$.

Taken together, the LC layer acts as a voltage-controlled phase element, allowing the selection of which diffraction order dominates the far-field emission. This voltage-tunable routing among discrete angular channels is a key feature of the single-mode device. Physically, the asymmetric in-plane bias creates a lateral gradient in the LC director field and therefore a spatially varying effective index for the emitted $x$ polarized field. The accumulated phase after propagation through an LC thickness $t_{\mathrm{LC}}$ is
\[
\phi(x,V)=\frac{2\pi}{\lambda}\int_{0}^{t_{\mathrm{LC}}} n_{\mathrm{eff}}(x,z;V)\,dz,
\]
where $n_{\mathrm{eff}}(x,z;V)$ is the local, voltage dependent effective index and $\lambda$ is the lasing wavelength. A nonzero lateral derivative, $\partial\phi/\partial x \neq 0$, represents a voltage-induced phase ramp across the emitting aperture.

According to the principles of Fourier optics, the angular intensity distribution in the far field is determined by the spatial variation of both amplitude and phase at the emitting surface. When the LC layer exhibits a nearly uniform phase distribution at low bias ($V<2$ V), the outgoing beam remains centered in the zeroth diffraction order. As the applied voltage increases, the LC behaves increasingly like a voltage-tunable phase grating. The phase increment between adjacent unit cells of the NHA, denoted by \(\Delta\phi(V)\), increases with \(V\) and dictates how optical power is redistributed among the discrete diffraction orders. When this phase increment approaches \(2\pi\), constructive interference favors the first diffraction order (around \(V \simeq 3~\mathrm{V}\)), while for \(\Delta\phi(V)\simeq4\pi\), the second order becomes dominant (around \(V \simeq 4.5~\mathrm{V}\)). Reversing the polarity of the applied bias reverses the sign of the lateral phase gradient, thereby flipping the beam deflection angle. In short, the applied voltage reorients the LC molecules, modulating the local effective index $n_{\mathrm{eff}}(x,z;V)$ and creating a tunable phase profile $\phi(x,V)$ across the aperture. This LC-induced phase grating controls the distribution of optical power among the diffraction orders, with certain bias points maximizing a particular diffraction order, while intermediate voltages can populate several orders with comparable strength.

In the periodic NHA, the allowed emission directions are constrained by the grating condition \(\sin\theta_m = m\lambda/\Lambda_{\mathrm{NHA}}\), where \(\Lambda_{\mathrm{NHA}}\) is the grating period and \(m\) is the diffraction order. In this device, the observed off-normal angles of $\pm 27^{\circ}$ and $\pm 67^{\circ}$ are simply the first and second diffraction orders of the NHA grating at $\lambda \approx 870~\text{nm}$. 
Therefore, the maximum steering range is set by the grating period and the wavelength. 
The LC birefringence only changes the strength of these orders and the voltages at which they become dominant; it does not alter their angular positions. Thus, the LC primarily modifies the phase profile. As a result, the cavity resonance and linewidth remain essentially unchanged. The spectral purity and beam divergence of the laser are also preserved.

Because each LC orientation corresponds to a distinct emission angle, the beam reconfiguration speed is governed by the LC response time. For a 3-$\mu$m-thick 5CB layer, a characteristic director reorientation
time scales as
$\tau \sim \gamma_{1} d^{2} / (\pi^{2} K_{11})$~\cite{lee2017two,wu1989design},
where $\gamma_{1}$ is the rotational viscosity of 5CB, $d$ is the LC layer
thickness, and $K_{11}$ is the splay elastic constant (Table~\ref{tab:5cb}).
The rotational viscosity $\gamma_{1}$ depends on temperature, material composition,
and the nematic order parameter. Here, we use a typical room-temperature value $\gamma_{1} \sim 0.1~\mathrm{Pa\cdot s}$ for 5CB reported in the literature
\cite{de1993physics}. Using these material parameters, this calculation yields a response time on the order of $10~\mathrm{ms}$ for our geometry.
In contrast, Geis \textit{et al.} reported 30--80~ns response in vertically driven cells under strong overdrive fields~\cite{Geis:10}, indicating that much faster LC dynamics are possible under large applied fields. These considerations suggest that LC dynamics, rather than the nanolaser cavity, currently set the limit on steering speed, and that thinner cells or stronger field configurations could significantly reduce the response time in future designs.

\subsection{Effect of LC Thickness and Periodicity}

The steering response of the LC-integrated PNL is strongly dependent on two structural parameters: the periodicity of the device and the thickness of the LC layer. Together, these factors determine both the number of available emission channels and the quality of angular control. To investigate the effect of periodicity, we varied the number of unit cells ($N$) on which a single LC segment was placed. The effective length of the LC region follows $L = (2N + 1)\Lambda_{1}$, where $\Lambda_{1}$ denotes the primitive lattice constant, so an increase in $N$ enlarges the overall structural period of the integrated device.

%----------------------------
% Figure 4
\begin{figure}[h]
\centering
\includegraphics[width=\linewidth, clip=true, trim=0cm 34.5cm 0cm 0cm]{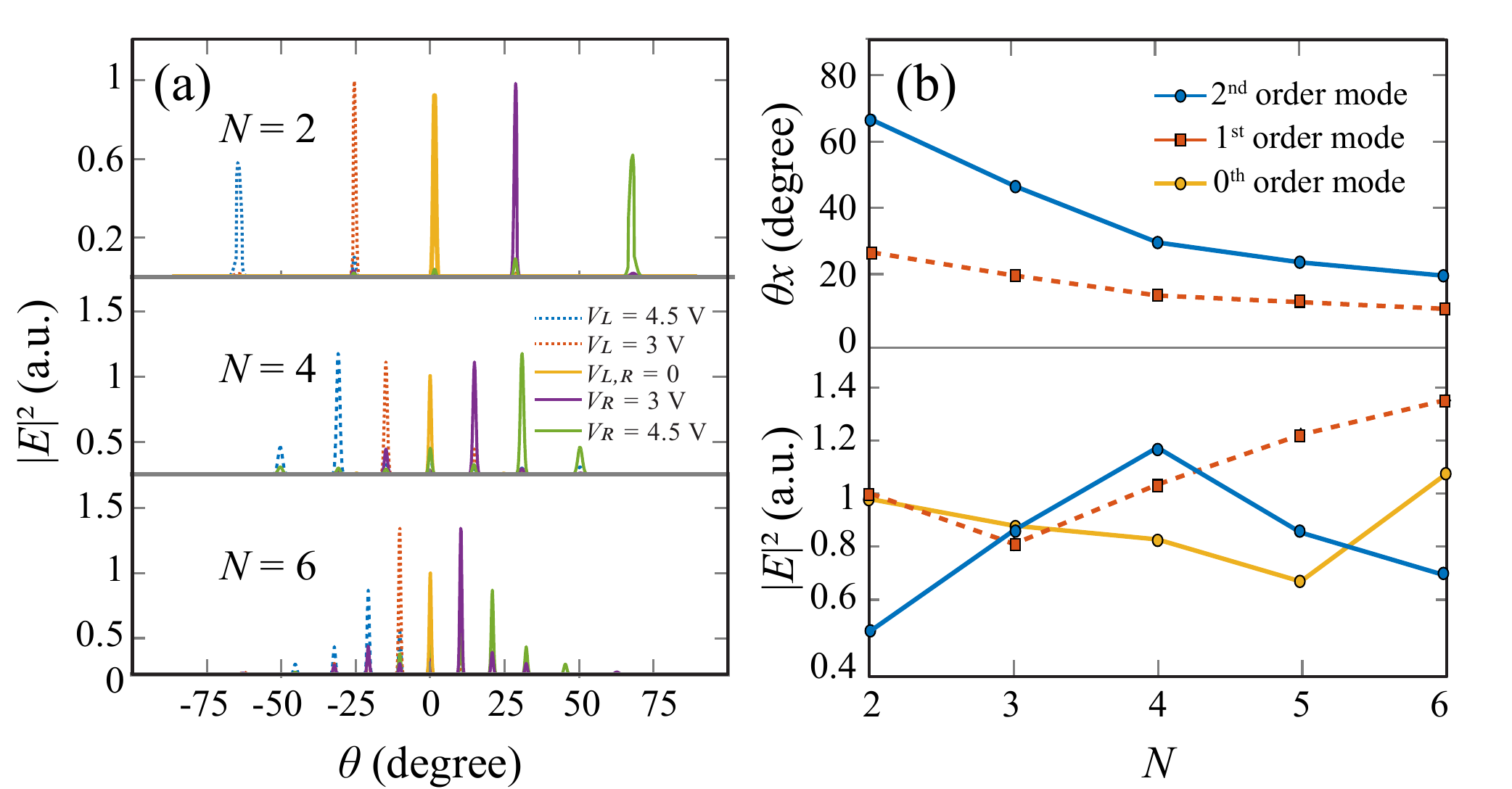}
\caption{Effect of periodicity on beam steering. 
(a) Far-field intensity distributions for $N=4$, $5$, and $6$, showing the emergence of additional diffraction orders as $N$ increases. 
(b) Angular position and corresponding far-field intensity of the $0^{\text{th}}$, $1^{\text{st}}$, and $2^{\text{nd}}$ order diffraction modes as a function of $N$.}
\label{fig:periodicity}
\end{figure}
%----------------------------

%---------------------------------
% Figure 5
\begin{figure}[t]
\centering
\includegraphics[width=\linewidth, clip=true, trim=0cm 30.5cm 1cm 0cm]{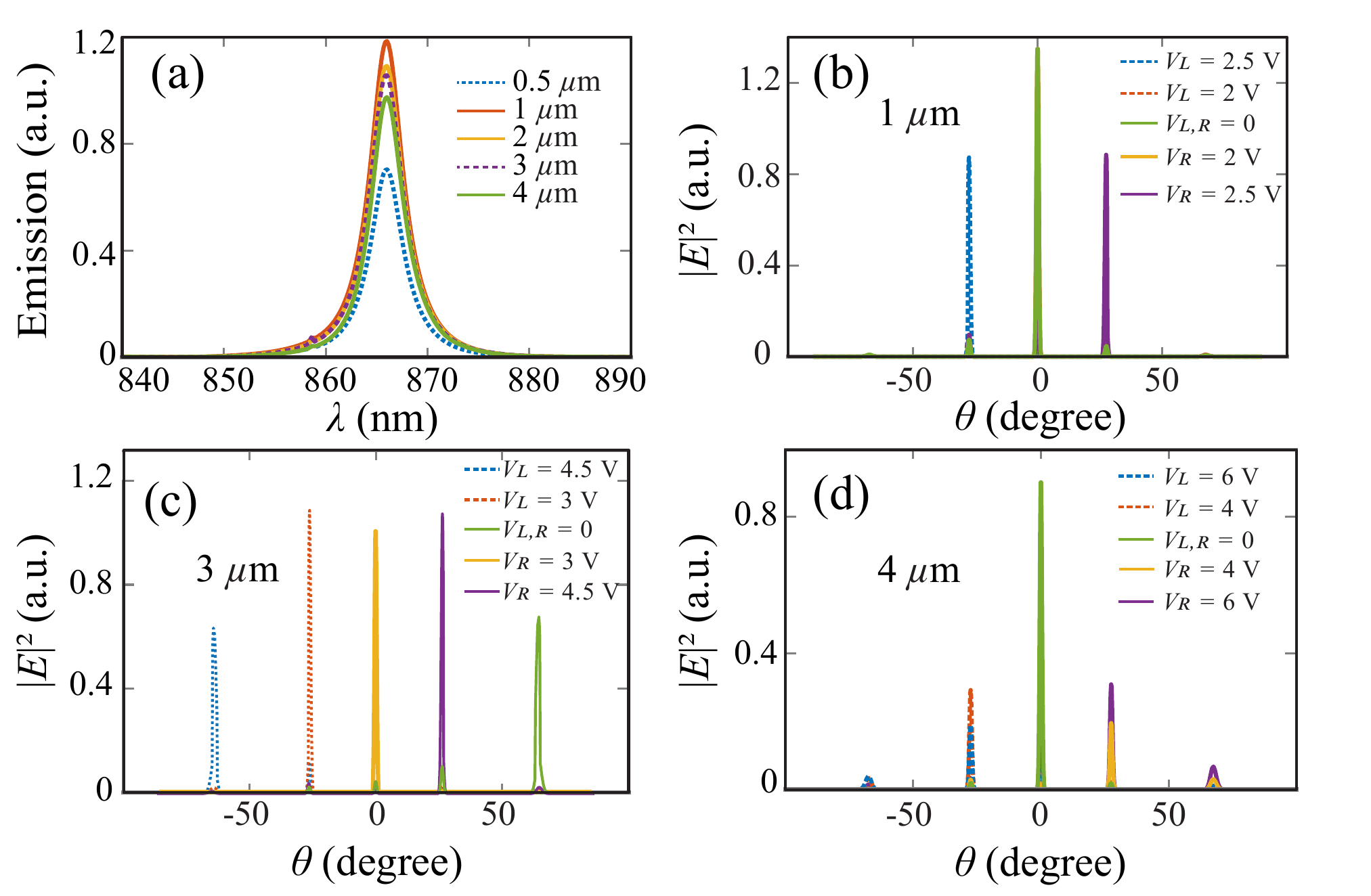}
\caption{Effect of LC thickness on emission and steering. 
(a) Emission spectra for LC thicknesses between 0.5--4 $\mu$m when both electrodes are grounded ($V_{L,R}=0$). 
(b--d) Far-field intensity distributions for thicknesses of 1 $\mu$m, 3 $\mu$m, and 4 $\mu$m, respectively. 
In each case, a voltage is applied to one electrode while the other is grounded ($V_{L}\neq0$, $V_{R}=0$ or $V_{R}\neq0$, $V_{L}=0$), with the metal NHA kept at 0~V. 
Optimum performance is observed at 3 $\mu$m, where emission intensity and angular selectivity are maximized.}
\label{fig:thickness}
\end{figure}
%----------------------------------------

As shown in Fig.~\ref{fig:periodicity}(a), the far-field profiles indicate that as $N$ increases, additional emission peaks appear, signifying the formation of higher-order diffraction modes, while the angular spacing between neighboring peaks becomes smaller. Figure \ref{fig:periodicity}(b) presents the variation of both the angular positions and the corresponding far-field intensities of the $0^{\text{th}}$, $1^{\text{st}}$, and $2^{\text{nd}}$ order modes as a function of $N$. The angular position of each order follows the expected trend, where higher orders appear at larger angles that gradually reduce with increasing periodicity. 

In contrast, the intensity variation does not follow a fixed monotonic pattern. The redistribution of energy among different diffraction orders arises from the combined influence of near-field interference and voltage-induced phase modulation within the LC layer. These effects occasionally enhance or suppress certain orders, particularly around intermediate values of $N$. The results also show that by adjusting the applied voltage, the relative contribution of each order can be tuned, although higher-order components beyond the second order remain weak because the LC orientation is not perfectly uniform across the device. The corresponding emission angles under different applied voltages are summarized in Table~\ref{tab:steering_angles}. All the results shown in Fig.~\ref{fig:periodicity}(a) and (b) correspond to a device dimension of $37.5~\mu\text{m} \times 37.5~\mu\text{m}$ and a pump field intensity of $1.8\times10^{8}~\text{V/m}$. Overall, increasing $N$ provides finer angular resolution and a greater number of steering states, while maintaining stable beam directionality when the LC thickness is kept at $3~\mu\text{m}$, which was identified as the optimal configuration for efficient steering.

The effect of LC thickness was studied separately for the case of $N=2$. Figure~\ref{fig:thickness}(a) shows that lasing is consistently observed at $\sim$870~nm across all thicknesses considered, confirming robust cavity feedback. Both the emission strength and the steering range are controlled by LC thickness. At $H=1~\mu\mathrm{m}$ the device delivers the highest intensity, yet steering is limited. The beam occupies only three angular channels, as shown in Fig.~\ref{fig:thickness}(b). The small thickness yields insufficient phase accumulation, which constrains tunability. Increasing to $H=3~\mu\mathrm{m}$ expands the accessible set to five well separated directions with narrow beam widths while maintaining high emission intensity, as shown in  Fig.~\ref{fig:thickness}(c). For even thicker cells, such as $H=4~\mu$m, beam steering becomes less efficient, as shown in Fig.~\ref{fig:thickness}(d). Although higher voltages can reorient the LC molecules, the intensity of higher-order modes remains suppressed compared to the central lobe, indicating excess phase retardation in the LC bulk.  

% Figure 6
These results establish that an LC thickness of $3~\mu\mathrm{m}$ provides the best compromise between emission strength and angular tunability, enabling robust steering into multiple diffraction orders with minimal degradation of beam quality. Periodicity, on the other hand, determines the angular separation and density of available emission states. Taken together, these findings confirm that both parameters must be co-optimized to realize efficient, voltage-controlled beam steering in LC-integrated PNLs.  
%----------------------------
% Table 3
\begin{table}[h]
\centering
\caption{Measured far-field emission angles (in degrees) for different periodicities ($N$) under various applied voltages.
A positive voltage on the left electrode ($V_{L}$) produces a negative steering angle, while a positive voltage on the right electrode ($V_{R}$) produces a positive steering angle.
The metal NHA was kept at $0$ V during all measurements, and all data correspond to an LC thickness of 3 µm.
All angles are rounded to the nearest integer.}
\renewcommand{\arraystretch}{1.35} % increased overall row height
\setlength{\tabcolsep}{7pt}        % spacing between columns
\begin{tabular}{|c|c|c|c|c|c|}
\hline
Periodicity ($N$) &
\shortstack{\rule{0pt}{3.5ex}\(V_{L}=4.5~\text{V}\)\\\(V_{R}=0~\text{V}\)} &
\shortstack{\rule{0pt}{3.5ex}\(V_{L}=3~\text{V}\)\\\(V_{R}=0~\text{V}\)} &
\shortstack{\rule{0pt}{3.5ex}\(V_{L}=0~\text{V}\)\\\(V_{R}=0~\text{V}\)} &
\shortstack{\rule{0pt}{3.5ex}\(V_{L}=0~\text{V}\)\\\(V_{R}=3~\text{V}\)} &
\shortstack{\rule{0pt}{3.5ex}\(V_{L}=0~\text{V}\)\\\(V_{R}=4.5~\text{V}\)} \\
\hline
2 & $-67^{\circ}$ & $-27^{\circ}$ & $0^{\circ}$ & $27^{\circ}$ & $67^{\circ}$ \\ \hline
3 & $-47^{\circ}$ & $-20^{\circ}$ & $0^{\circ}$ & $20^{\circ}$ & $47^{\circ}$ \\ \hline
4 & $-31^{\circ}$ & $-15^{\circ}$ & $0^{\circ}$ & $15^{\circ}$ & $31^{\circ}$ \\ \hline
5 & $-25^{\circ}$ & $-12^{\circ}$ & $0^{\circ}$ & $12^{\circ}$ & $25^{\circ}$ \\ \hline
6 & $-21^{\circ}$ & $-10^{\circ}$ & $0^{\circ}$ & $10^{\circ}$ & $21^{\circ}$ \\ \hline
\end{tabular}
\label{tab:steering_angles}
\end{table}
%---------------------------------
\subsection{Dual-Mode Nanolaser: Beam Steering}

To extend steering capability to multiple wavelengths, the standard NHA was replaced with a merged lattice, as shown in Fig.~\ref{fig:1}(c), while keeping the remainder of the stack unchanged. The merged array supports two distinct resonant modes within the same cavity, enabling dual-mode lasing. The configuration analyzed here corresponds to $N=3$, where a single LC segment spans three unit cells along the $x$ direction. The device is periodic in both $x$ and $y$, and is pumped by an $x$-polarized plane wave incident along the $+z$ direction at $\lambda=800$\,nm. A reduced pump amplitude of $7\times10^{7}$\,V/m is sufficient for lasing in this geometry, consistent with the lower threshold typically observed in merged NHAs compared to simple square lattices \cite{D1NA00402F}. Far-field responses were calculated for a device area of $45~\mu$m $\times$ $45~\mu$m.  

%----------------------------
% Figure 6
\begin{figure*}[htbp]
\centering
\includegraphics[width=0.9\textwidth, clip=true, trim=0cm 29.5cm 3cm 0.75cm]{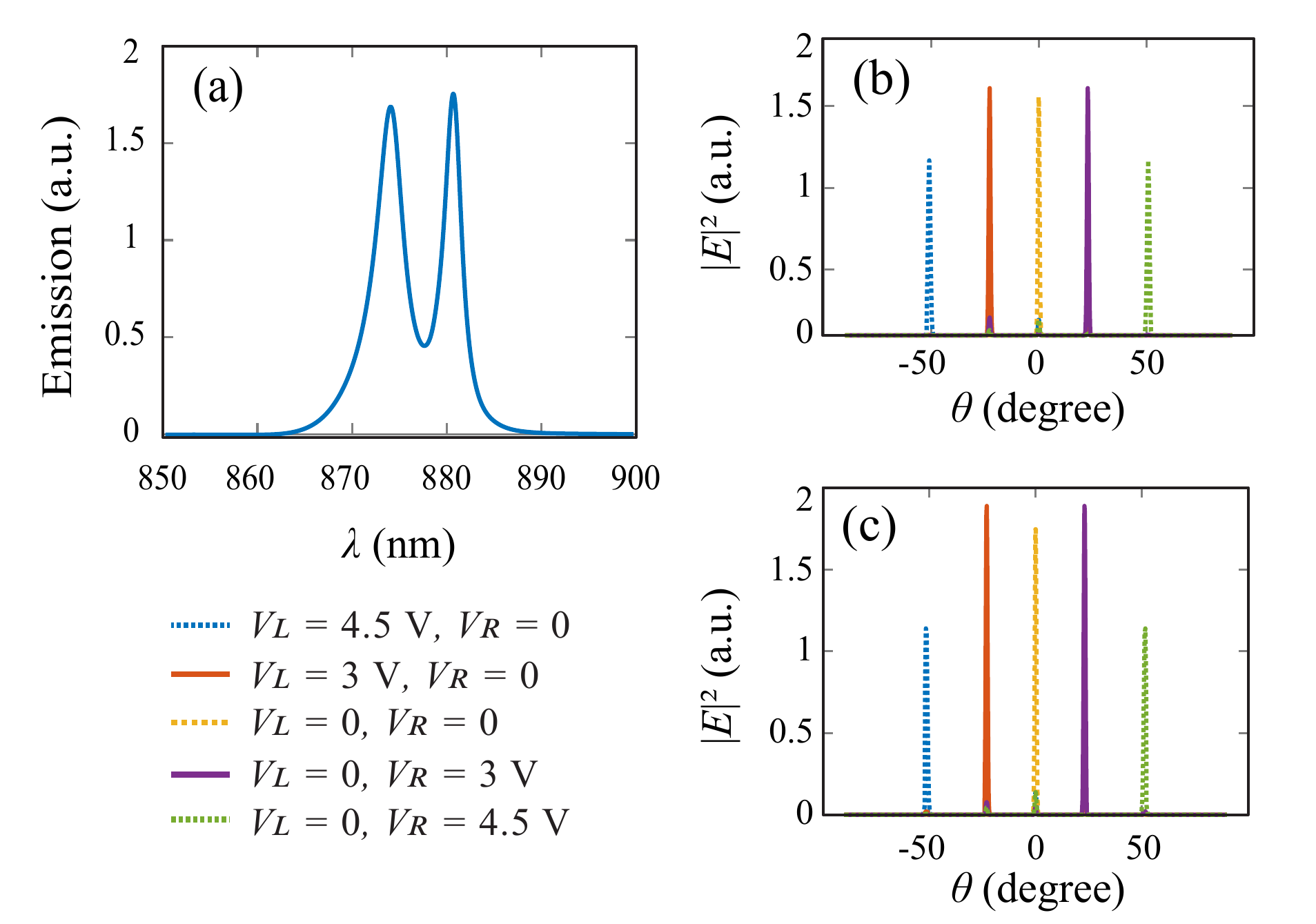}
\caption{Dual-mode LC-integrated plasmonic nanolaser with a merged NHA for $N=3$. (a) Emission spectrum showing two narrow lasing peaks at 873 nm and 880 nm, confirming dual-mode operation. (b) Far-field intensity distributions of the 873 nm mode for different applied voltages, demonstrating discrete beam steering at $-51^{\circ}$, $-23^{\circ}$, $0^{\circ}$, $23^{\circ}$, and $51^{\circ}$. (c) Corresponding far-field patterns for the 880 nm mode, showing similar angular trajectories with slightly higher output intensity. 
All results correspond to a 3 $\mu$m LC thickness.}
\label{fig:dual_mode}
\end{figure*}
%------------------------------

The emission spectrum in Fig.~\ref{fig:dual_mode} (a) shows two narrow peaks at 873\,nm and 880\,nm with linewidths below 3\,nm. The spectrum indicates dual-mode lasing, with any broadening attributable to the LC layer appearing to be small. The merged lattice supports closely spaced plasmonic resonances, so two modes can appear in the same cavity at slightly different wavelengths. The separation is about 7\,nm, small enough to indicate a shared cavity yet large enough to address the lines by wavelength when needed.

Far field intensity distributions in Fig.~\ref{fig:dual_mode} (b) and (c) show how each mode steers with applied voltage. In both cases, the emission occupies five angular states at $-51^{\circ}$, $-23^{\circ}$, $0^{\circ}$, $23^{\circ}$, and $51^{\circ}$. These angles are consistent with diffraction orders set by the LC-mediated phase gradient. The two modes follow almost the same trajectory, and their far-field patterns overlap within the resolution of the monitor. This overlap means a single routing path can serve either line, which is useful for integration.

At the pump level used here, the 880\,nm line gives a higher output than the 873\,nm line. The difference may reflect slightly better coupling between the phase profile and the cavity at 880\,nm. Even so, both modes keep narrow angular widths of about $1^{\circ}$, so beam quality appears unchanged during two-color operation. Within the tested range, the voltage-driven tuning was reversible: angles and spectra returned to their initial values after several cycles. We did not see measurable drift in linewidth or peak position during steering, although longer runs would be needed to rule out slow changes.

Taken together, the data indicate that the merged lattice nanolaser supports voltage-controlled steering at two wavelengths while maintaining coherence and directivity within the limits of our measurement. The same steering of two modes could enable dual-channel links, simple wavelength division multiplexing, or multi-target sensing in settings that favor compact and reconfigurable sources.

\subsection{Comparative Analysis and Applications}

The results in the preceding sections demonstrate that an LC layer directs PNL emission through a voltage-controlled phase gradient. In the single-mode case, the beam shifts by redistributing power among discrete diffraction orders. By selecting the driven electrode and setting the bias, the output can be placed in specific angular channels while the angular width remains near $1^{\circ}$. A thickness near $H\approx 3~\mu\mathrm{m}$ provides a practical balance between emitted power and tuning range. The periodicity parameter $N$ sets the count and spacing of the available steering states.

With a merged lattice NHA, the device supports two lasing modes at $873$ and $880\,\mathrm{nm}$. Both lasing modes were steered to five angles ($-51^{\circ}$, $-23^{\circ}$, $0^{\circ}$, $23^{\circ}$, $51^{\circ}$) under voltage control, and their far field trajectories nearly overlapped. Under the same pump, the $880~\mathrm{nm}$ mode displayed stronger emission, but both maintained narrow angular widths of about $1^{\circ}$. Compared to the single-mode case, the dual-mode nanolaser offers the additional advantage of delivering two coherent channels through the same steering mechanism without compromising beam quality.

These results suggest several application directions. In integrated photonics, voltage-controlled steering can replace moving elements for routing on-chip signals \cite{miller2009device}. The dual-mode behavior appears suitable for wavelength division multiplexing \cite{hill2014advances}, where two data streams share one routing path. Steering among discrete angles also matches tasks in nanoscale scanning, spatial imaging, and multi-sensor readout \cite{gao2014single}. Related ideas include beam division multiplexing, optical trapping, and compact on-chip LiDAR \cite{Poulton:17}.

Overall, the comparative study underscores the versatility of LC integration with PNLs: single-mode devices provide precise and controllable beam redirection, while dual-mode devices extend this capability to multi-wavelength operation. This dual functionality establishes LC integrated PNLs as promising candidates for compact, reconfigurable, and multifunctional nanophotonic platforms.

\section{Conclusion}
We have demonstrated that integrating an LC layer above a PNL enables electrically programmable beam steering without altering the cavity geometry. Numerical simulations coupling LC director reorientation with full-vector FDTD modeling reveal that voltage-induced phase gradients across the LC layer redistribute power among the grating diffraction orders, steering the output into discrete angular states while preserving beam quality and spectral purity. For the single-mode device, the steering angle is set primarily by the grating period, whereas the LC voltage determines which diffraction order dominates. An LC thickness of approximately 3 $\mu$m offers the best balance between intensity and angular tunability, and the periodicity parameter $N$ controls the number and spacing of accessible steering angles.

Extending the design to a merged nanohole lattice enables dual-mode lasing at 873 nm and 880 nm. Both modes follow nearly identical voltage-dependent trajectories, resulting in stable and reversible steering to five distinct angular states. The dual-mode operation preserves coherence and beam divergence, demonstrating that a single cavity can support two independently usable optical channels. Overall, LC integration transforms PNLs from fixed-geometry emitters into voltage-reconfigurable beam sources, making them suitable for applications in on-chip routing, multiplexed communication, nanoscale imaging, and adaptive photonic interfaces. %Future work will focus on experimental realization, optimization of LC switching dynamics, and strategies for independently steering multiple lasing lines within the same device.

\section*{Disclosures}
% \section*{Disclosures}
The authors declare no conflicts of interest.

\bibliography{sample}

@article{zia2006plasmonics,
  title={Plasmonics: the next chip-scale technology},
  author={Zia, Rashid and Schuller, Jon A and Chandran, Anu and Brongersma, Mark L},
  journal={Materials today},
  volume={9},
  number={7-8},
  pages={20--27},
  year={2006},
  publisher={Elsevier}
}

@article{pleros2011optical,
  title={Optical interconnects using plasmonics and Si-photonics},
  author={Pleros, Nikos and Kriezis, Emmanouil E and Vyrsokinos, Konstantinos},
  journal={IEEE Photonics Journal},
  volume={3},
  number={2},
  pages={296--301},
  year={2011},
  publisher={IEEE}
}

@article{ohashi2009chip,
  title={On-chip optical interconnect},
  author={Ohashi, Keishi and Nishi, Kenichi and Shimizu, Takanori and Nakada, Masafumi and Fujikata, Junichi and Ushida, Jun and Torii, Sunao and Nose, Koichi and Mizuno, Masayuki and Yukawa, Hiroaki and others},
  journal={Proceedings of the IEEE},
  volume={97},
  number={7},
  pages={1186--1198},
  year={2009},
  publisher={IEEE}
}

@article{stucchi2013chip,
  title={On-chip optical interconnects versus electrical interconnects for high-performance applications},
  author={Stucchi, Michele and Cosemans, Stefan and Van Campenhout, Joris and T{\H{o}}kei, Zsolt and Beyer, Gerald},
  journal={Microelectronic engineering},
  volume={112},
  pages={84--91},
  year={2013},
  publisher={Elsevier}
}

@article{koo2002bragg,
  title={Bragg grating-based laser sensors systems with interferometric interrogation and wavelength division multiplexing},
  author={Koo, KP and Kersey, AD},
  journal={Journal of Lightwave Technology},
  volume={13},
  number={7},
  pages={1243--1249},
  year={2002},
  publisher={IEEE}
}

@article{ozcelik2015optofluidic,
  title={Optofluidic wavelength division multiplexing for single-virus detection},
  author={Ozcelik, Damla and Parks, Joshua W and Wall, Thomas A and Stott, Matthew A and Cai, Hong and Parks, Joseph W and Hawkins, Aaron R and Schmidt, Holger},
  journal={Proceedings of the National Academy of Sciences},
  volume={112},
  number={42},
  pages={12933--12937},
  year={2015},
  publisher={National Academy of Sciences}
}

@article{momtaj2025toward,
  title={Toward low-noise on-chip plasmonic three-dimensional biological cell imaging},
  author={Momtaj, Maliha and Talukder, Muhammad Anisuzzaman},
  journal={Optics Express},
  volume={33},
  number={3},
  pages={3901--3919},
  year={2025},
  publisher={Optica Publishing Group}
}

@article{bergman2003surface,
  title={Surface Plasmon Amplification by Stimulated Emission of Radiation: Quantum Generation of Coherent Surface Plasmons in Nanosystems},
  author={Bergman, David J and Stockman, Mark I},
  journal={Physical review letters},
  volume={90},
  number={2},
  pages={027402},
  year={2003},
  publisher={APS}
}

@article{hill2007lasing,
  title={Lasing in metallic-coated nanocavities},
  author={Hill, Martin T and Oei, Yok-Siang and Smalbrugge, Barry and Zhu, Youcai and De Vries, Tjibbe and Van Veldhoven, Peter J and Van Otten, Frank WM and Eijkemans, Tom J and Turkiewicz, Jaros{\l}aw P and De Waardt, Huug and others},
  journal={Nature Photonics},
  volume={1},
  number={10},
  pages={589--594},
  year={2007},
  publisher={Nature Publishing Group UK London}
}

@article{ma2011room,
  title={Room-temperature sub-diffraction-limited plasmon laser by total internal reflection},
  author={Ma, Ren-Min and Oulton, Rupert F and Sorger, Volker J and Bartal, Guy and Zhang, Xiang},
  journal={Nature materials},
  volume={10},
  number={2},
  pages={110--113},
  year={2011},
  publisher={Nature Publishing Group UK London}
}

@article{noginov2009demonstration,
  title={Demonstration of a spaser-based nanolaser},
  author={Noginov, MA and Zhu, G and Belgrave, AM and Bakker, Reuben and Shalaev, VM and Narimanov, EE and Stout, S and Herz, E and Suteewong, T and Wiesner, U},
  journal={Nature},
  volume={460},
  number={7259},
  pages={1110--1112},
  year={2009},
  publisher={Nature Publishing Group UK London}
}

@article{lu2012plasmonic,
  title={Plasmonic nanolaser using epitaxially grown silver film},
  author={Lu, Yu-Jung and Kim, Jisun and Chen, Hung-Ying and Wu, Chihhui and Dabidian, Nima and Sanders, Charlotte E and Wang, Chun-Yuan and Lu, Ming-Yen and Li, Bo-Hong and Qiu, Xianggang and others},
  journal={science},
  volume={337},
  number={6093},
  pages={450--453},
  year={2012},
  publisher={American Association for the Advancement of Science}
}

@article{khajavikhan2012thresholdless,
  title={Thresholdless nanoscale coaxial lasers},
  author={Khajavikhan, M and Simic, A and Katz, M and Lee, JH and Slutsky, B and Mizrahi, A and Lomakin, V and Fainman, Y},
  journal={Nature},
  volume={482},
  number={7384},
  pages={204--207},
  year={2012},
  publisher={Nature Publishing Group UK London}
}

@article{deeb2017plasmon,
  title={Plasmon lasers: coherent nanoscopic light sources},
  author={Deeb, Claire and Pelouard, Jean-Luc},
  journal={Physical Chemistry Chemical Physics},
  volume={19},
  number={44},
  pages={29731--29741},
  year={2017},
  publisher={Royal Society of Chemistry}
}

@article{zhang2014room,
  title={A room temperature low-threshold ultraviolet plasmonic nanolaser},
  author={Zhang, Qing and Li, Guangyuan and Liu, Xinfeng and Qian, Fang and Li, Yat and Sum, Tze Chien and Lieber, Charles M and Xiong, Qihua},
  journal={Nature communications},
  volume={5},
  number={1},
  pages={4953},
  year={2014},
  publisher={Nature Publishing Group UK London}
}

@Article{D1NA00402F,
author ="Shahid, Shadman and Zumrat, Shahed-E- and Talukder, Muhammad Anisuzzaman",
title  ="A merged lattice metal nanohole array based dual-mode plasmonic laser with an ultra-low threshold",
journal  ="Nanoscale Adv.",
year  ="2022",
volume  ="4",
issue  ="3",
pages  ="801-813",
publisher  ="RSC",
doi  ="10.1039/D1NA00402F",
url  ="http://dx.doi.org/10.1039/D1NA00402F"}

@article{zumrat2022dual,
  title={Dual-wavelength hybrid Tamm plasmonic laser},
  author={Zumrat, Shahed-E- and Shahid, Shadman and Talukder, Muhammad Anisuzzaman},
  journal={Optics Express},
  volume={30},
  number={14},
  pages={25234--25248},
  year={2022},
  publisher={Optica Publishing Group}
}

@book{de1993physics,
  title={The physics of liquid crystals},
  author={De Gennes, Pierre-Gilles and Prost, Jacques},
  number={83},
  year={1993},
  publisher={Oxford university press}
}

@article{wu1989design,
  title={Design of a liquid crystal based tunable electrooptic filter},
  author={Wu, Shin-Tson},
  journal={Applied optics},
  volume={28},
  number={1},
  pages={48--52},
  year={1989},
  publisher={Optical Society of America}
}

@article{lee2017two,
  title={Two-photon polymerization enabled multi-layer liquid crystal phase modulator},
  author={Lee, Yun-Han and Franklin, Daniel and Gou, Fangwang and Liu, Guigeng and Peng, Fenglin and Chanda, Debashis and Wu, Shin-Tson},
  journal={Scientific Reports},
  volume={7},
  number={1},
  pages={16260},
  year={2017},
  publisher={Nature Publishing Group UK London}
}

@article{frank1958liquid,
  title={I. Liquid crystals. On the theory of liquid crystals},
  author={Frank, Frederick C},
  journal={Discussions of the Faraday Society},
  volume={25},
  pages={19--28},
  year={1958},
  publisher={Royal Society of Chemistry}
}

@misc{COMSOL62,
  author       = {{COMSOL AB}},
  title        = {{COMSOL Multiphysics} v. 6.2},
  howpublished = {\url{https://www.comsol.com}},
  address      = {Stockholm, Sweden},
  year         = {2023},
  note         = {Finite element and multiphysics simulation software}
}

@article{azad2021mode,
  title={Mode-resolved analysis of a planar multi-layer plasmonic nanolaser},
  author={Azad, Zihad and Talukder, Muhammad Anisuzzaman},
  journal={Optics Communications},
  volume={482},
  pages={126614},
  year={2021},
  publisher={Elsevier}
}

@article{azad2022,
  title={Simultaneously surface- and edge-emitting plasmonic laser operating in the near-infrared region},
  author={Azad, Zihad and Islam, Md Shofiqul and Talukder, Muhammad Anisuzzaman},
  journal={Optics \& Laser Technology},
  volume={146},
  pages={107571},
  year={2022},
  publisher={Elsevier}
}

@article{ahmed2018efficient,
  title={An efficient and directional optical Tamm state assisted plasmonic nanolaser with broad tuning range},
  author={Ahmed, Zabir and Talukder, Muhammad Anisuzzaman},
  journal={Journal of Physics Communications},
  volume={2},
  number={4},
  pages={045016},
  year={2018},
  publisher={IOP Publishing}
}

@article{shahid2025beyond,
  title={Beyond periodicity: tailoring Tamm resonances in plasmonic nanohole arrays for multimodal lasing},
  author={Shahid, Shadman and Talukder, Muhammad Anisuzzaman},
  journal={New Journal of Physics},
  volume={27},
  number={1},
  pages={013014},
  year={2025},
  publisher={IOP Publishing}
}

@article{ahamed2024wavelength,
  title={Wavelength selective beam-steering in a dual-mode multi-layer plasmonic laser},
  author={Ahamed, Mahin and Afroj, Md Nasim and Shahid, Shadman and Talukder, Muhammad Anisuzzaman},
  journal={Optics Express},
  volume={32},
  number={11},
  pages={19895--19909},
  year={2024},
  publisher={Optica Publishing Group}
}

@article{Winkler2020DualWavelengthLI,
  title={Dual-Wavelength Lasing in Quantum-Dot Plasmonic Lattice Lasers.},
  author={Jan M Winkler and Max Josef Ruckriegel and Henar Rojo and Robert C. Keitel and Eva De Leo and Freddy T. Rabouw and David J. Norris},
  journal={ACS nano},
  year={2020},
  url={https://api.semanticscholar.org/CorpusID:210839361}
}

@book{yang2014fundamentals,
  title={Fundamentals of liquid crystal devices},
  author={Yang, Deng-Ke and Wu, Shin-Tson},
  year={2014},
  publisher={John Wiley \& Sons}
}

@article{bahadur1990dynamic,
  title={Dynamic scattering mode LCDs},
  author={Bahadur, Birendra},
  journal={Liquid Crystals Applications and Uses},
  volume={1},
  year={1990}
}

@article{lu2005ultrawide,
  title={Ultrawide-view liquid crystal displays},
  author={Lu, Ruibo and Zhu, Xinyu and Wu, Shin-Tson and Hong, Qi and Wu, Thomas X},
  journal={Journal of Display Technology},
  volume={1},
  number={1},
  pages={3},
  year={2005},
  publisher={OSA}
}

@article{assanto2012nematicons,
  title={Nematicons, spatial optical solitons in nematic liqud crystals},
  author={Assanto, G},
  journal={John WileyandSons, NewYork},
  year={2012}
}

@inproceedings{sautter2015active,
  title={Active tuning of all-dielectric metasurfaces with liquid crystals},
  author={Sautter, J{\"u}rgen and Staude, Isabelle and Decker, Manuel and Rusak, Evgenia and Neshev, Dragomir N and Brener, Igal and Kivshar, Yuri S},
  booktitle={The European Conference on Lasers and Electro-Optics},
  pages={CK\_6\_3},
  year={2015},
  organization={Optica Publishing Group}
}

@phdthesis{komar2018tunable,
  title={Tunable All-dielectric Metasurfaces: Fundamentals and Applications},
  author={Komar, Andrei},
  year={2018},
  school={The Australian National University (Australia)}
}

@article{cirtoaje2020freedericksz,
  title={Freedericksz Transitions in Twisted Ferronematics Subjected to Magnetic and Laser Field},
  author={Cirtoaje, Cristina and Iacobescu, Gabriela and Petrescu, Emil},
  journal={Crystals},
  volume={10},
  number={7},
  pages={567},
  year={2020},
  publisher={MDPI}
}

@article{yee1966numerical,
  title={Numerical solution of initial boundary value problems involving Maxwell's equations in isotropic media},
  author={Yee, Kane},
  journal={IEEE Transactions on antennas and propagation},
  volume={14},
  number={3},
  pages={302--307},
  year={1966},
  publisher={Ieee}
}

@misc{lumericalFDTD,
  author       = {{Ansys Lumerical Inc.}},
  title= {Lumerical FDTD Solutions},
  howpublished = {\url{https://www.lumerical.com/products/fdtd/}},
  note         = {Accessed: 2025-09-13}
}

@article{dridi2013model,
  title={Model for describing plasmon-enhanced lasers that combines rate equations with finite-difference time-domain},
  author={Dridi, Montacer and Schatz, George C},
  journal={Journal of the Optical Society of America B},
  volume={30},
  number={11},
  pages={2791--2797},
  year={2013},
  publisher={Optical Society of America}
}

@article{chang2004finite,
  title={Finite-difference time-domain model of lasing action in a four-level two-electron atomic system},
  author={Chang, Shih-Hui and Taflove, Allen},
  journal={Optics express},
  volume={12},
  number={16},
  pages={3827--3833},
  year={2004},
  publisher={Optical Society of America}
}

@article{miller2009device,
  title={Device requirements for optical interconnects to silicon chips},
  author={Miller, David AB},
  journal={Proceedings of the IEEE},
  volume={97},
  number={7},
  pages={1166--1185},
  year={2009},
  publisher={IEEE}
}

@article{hill2014advances,
  title={Advances in small lasers},
  author={Hill, Martin T and Gather, Malte C},
  journal={Nature Photonics},
  volume={8},
  number={12},
  pages={908--918},
  year={2014},
  publisher={Nature Publishing Group UK London}
}

@article{gao2014single,
  title={Single-shot compressed ultrafast photography at one hundred billion frames per second},
  author={Gao, Liang and Liang, Jinyang and Li, Chiye and Wang, Lihong V},
  journal={Nature},
  volume={516},
  number={7529},
  pages={74--77},
  year={2014},
  publisher={Nature Publishing Group UK London}
}

@article{Geis:10,
author = {M. W. Geis and T. M. Lyszczarz and R. M. Osgood and B. R. Kimball},
journal = {Opt. Express},
number = {18},
pages = {18886--18893},
publisher = {Optica Publishing Group},
title = {30 to 50 ns liquid-crystal optical switches},
volume = {18},
month = {Aug},
year = {2010},
url = {https://opg.optica.org/oe/abstract.cfm?URI=oe-18-18-18886},
doi = {10.1364/OE.18.018886},
}

@article{mur2022controllable,
  title={Controllable shifting, steering, and expanding of light beam based on multi-layer liquid-crystal cells},
  author={Mur, Urban and Ravnik, Miha and Se{\v{c}}, David},
  journal={Scientific reports},
  volume={12},
  number={1},
  pages={352},
  year={2022},
  publisher={Nature Publishing Group UK London}
}

@article{Poulton:17,
author = {Christopher V. Poulton and Ami Yaacobi and David B. Cole and Matthew J. Byrd and Manan Raval and Diedrik Vermeulen and Michael R. Watts},
journal = {Opt. Lett.},
number = {20},
pages = {4091--4094},
publisher = {Optica Publishing Group},
title = {Coherent solid-state LIDAR with silicon photonic optical phased arrays},
volume = {42},
month = {Oct},
year = {2017},
url = {https://opg.optica.org/ol/abstract.cfm?URI=ol-42-20-4091},
doi = {10.1364/OL.42.004091},
}
\end{document}